\documentclass{vgtc}                          

\graphicspath{{figures/}{pictures/}{images/}{./}} 

\usepackage{times}                     

\usepackage{tabu}                      
\usepackage{booktabs}                  
\usepackage{multirow}
\usepackage{lipsum}                    
\usepackage{mwe}                       
\usepackage{url}
\usepackage{mathptmx}                  
\usepackage{makecell} 
\usepackage[utf8]{inputenc}
\usepackage{array}
\usepackage{caption}
\usepackage{amsmath}  
\usepackage{textalpha} 
\usepackage{xcolor}
\usepackage{setspace}

\usepackage{array}
\usepackage{geometry}
\newcolumntype{L}[1]{>{\raggedright\arraybackslash}p{#1}}
\newcolumntype{C}[1]{>{\centering\arraybackslash}p{#1}}
\newcolumntype{R}[1]{>{\raggedleft\arraybackslash}p{#1}}

\onlineid{2026}

\vgtccategory{Research}

\vgtcinsertpkg

\title{Bridging the Gap: Enhancing Gaze-Performance Link in Children with ASD through Dual-Level Visual Guidance in MR-DMT}

\author{Weiying Liu\\ %
        \scriptsize Shandong University %
\and Yanran Yuan\\ %
     \scriptsize Jining No.1 People's Hospital %
\and Zhiqiang Sheng\\ %
     \scriptsize Jining No.1 People's Hospital %
\and Dandan Lian\\ %
     \scriptsize Jining No.1 People's Hospital %
\and Sheng Li\thanks{e-mail: lisheng@pku.edu.cn}\\ %
     \scriptsize Jining No.1 People's Hospital %
\and Yufan Zhang\\ %
     \scriptsize Shandong University 
\and Yulong Bian\thanks{e-mail: ed.bianyulong@sdu.edu.cn}\\ %
     \scriptsize Shandong University 
\and Juan Liu\thanks{e-mail: zzzliujuan@sdu.edu.cn}\\ %
     \scriptsize Shandong University 
     }%

\teaser{
  \centering
  \includegraphics[width=\linewidth]{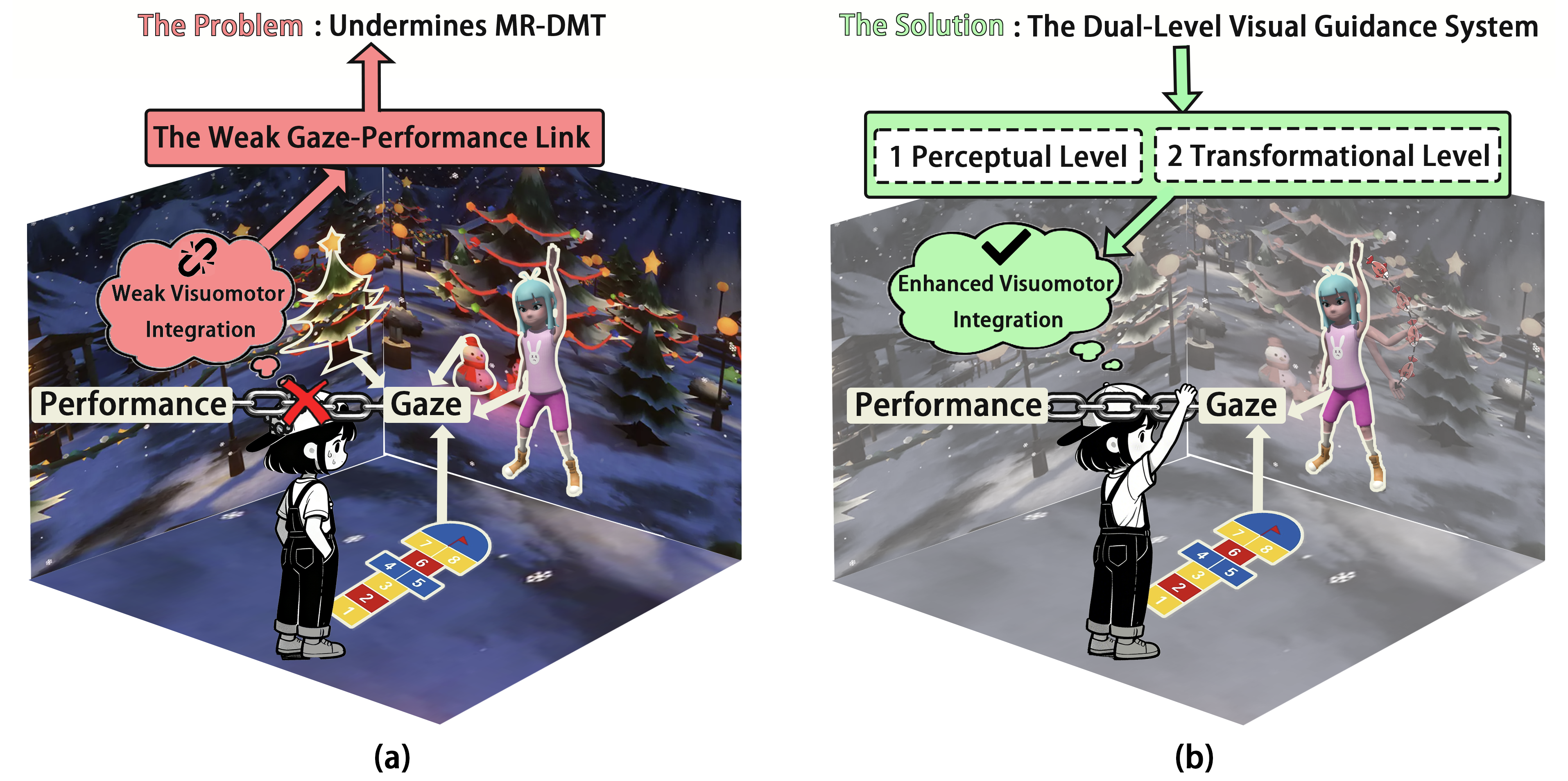}
  \caption{Visual Summary of the Gaze-Performance Link and Its Intervention.
\textbf{(a) The Problem:} In a baseline MR-DMT environment without guidance, children with ASD exhibit a weak gaze-performance link. Although visual attention (gaze trajectory) may be directed toward the virtual agent, this does not readily translate into accurate motor imitation (disrupted motor pathway), highlighting the core deficit in visuomotor integration (VMI). 
\textbf{(b) The Solution:} The proposed dual-level visual guidance system addresses this gap. Perceptual guidance (salience modulation) dims distractions and creates a spotlight effect on the agent. Transformational guidance (kinetic metaphors) provides predictive motor cues. Together, they strengthen the association, enabling gaze to effectively guide motor execution. }
  \label{fig:teaser}
}

\abstract{
   Autism Spectrum Disorder (ASD) is marked by action imitation deficits stemming from visuomotor integration impairments, posing challenges to imitation-based learning, such as dance movement therapy in mixed reality (MR-DMT). Previous gaze-guiding interventions in ASD have mainly focused on optimizing gaze in isolation, neglecting the crucial ``gaze-performance link''. This study investigates enhancing this link in MR-DMT for children with ASD. Initially, we experimentally confirmed the weak link: longer gaze durations didn't translate to better performance. Then, we proposed and validated a novel dual-level visual guidance system that operates on both perceptual and transformational levels: not only directing attention to task-relevant areas but also explicitly scaffolding the translation from gaze perception to performance execution. Our results demonstrate its effectiveness in boosting the gaze-performance link, laying key foundations for more precisely tailored and effective MR-DMT interventions for ASD.
} 


\begin{document}


\firstsection{Introduction}

\maketitle

Dance Movement Therapy (DMT) \cite{takahashi2019effectiveness, aithal2021dance} is a well-established intervention for Autism Spectrum Disorder (ASD) that uses movement to support the emotional, social, cognitive, and physical integration of individuals, to improve overall health and well-being \cite{ADTA2009}. However, traditional DMT approaches often face limitations in ecological validity and practical implementation. Mixed Reality (MR) environments offer a promising avenue to overcome these challenges, as studies have shown that children with ASD exhibit increased engagement in MR-based interventions~\cite{liu2022evaluating, liu2021designing}. In a recent study, Liu et al. \cite{liu2024self} developed a self-guided mixed reality dance movement therapy (MR-DMT) system for children with ASD, which uses a virtual agent to guide participants through the imitation of dance movements, resulting in significant performance improvements. Nonetheless, their study did not specifically examine how neurocognitive characteristics associated with ASD—particularly relevant given MR’s unique sensory demands—might influence therapeutic outcomes.

ASD is characterized by persistent deficits in social communication and interaction, along with restricted and repetitive behavioral patterns \cite{edition1980diagnostic}. A core challenge for children with ASD, especially in younger age groups, lies in action imitation~\cite{williams2004systematic}. Research suggests that this difficulty arises from impairments in visuomotor integration (VMI) mechanisms \cite{lidstone2021moving, williams2004systematic}, where the translation of observed actions into motor execution is disrupted. Furthermore, deficits in VMI have been extensively documented in individuals with ASD \cite{gepner2002rapid, miller2014dyspraxia}. Together, these neurocognitive challenges may compromise the effectiveness of MR-DMT, which relies on imitation-based learning and robust visuo-motor pathways. This gap highlights the need for innovative approaches that can specifically enhance the association between visual input and motor output in children with ASD.

MR technology presents a potential solution by enabling precisely controlled visual environments that can systematically modulate perceptual input. Unlike traditional DMT settings, MR allows for the implementation of targeted visual guidance strategies. Due to the atypical gaze patterns commonly observed in children with ASD \cite{nakano2010atypical, drysdale2018gaze, ishizaki2021eye}, numerous studies have explored methods to direct their gaze toward task-relevant areas \cite{wang2020promoting, liu2017feasibility, mcparland2021investigating, feng2013can, palestra2017artificial, shamir2023metacognitive, lee2022improving, lahiri2011dynamic}. However, many existing approaches focus on optimizing isolated metrics like gaze behavior, often overlooking the critical \textit{gaze-performance link}. Critically, the gaze-performance link serves as a direct behavioral proxy for assessing the functional integrity of the underlying VMI process. A strong link indicates that visual information is being efficiently converted into motor commands, which is the hallmark of effective VMI. Conversely, when extended gaze duration does not lead to corresponding improvements in performance, it indicates a weak gaze-performance link and a potential breakdown in VMI. In such cases, interventions focusing solely on guiding gaze may have limited efficacy. Thus, it is essential to focus on the visuomotor coupling function in children with ASD during MR-based DMT and to explore effective methods for enhancing this coupling. This critical research gap is the focus of our study.

The primary objective of this study is to investigate the existence of a weak gaze-performance link in children with ASD within MR-DMT and to propose methodological strategies to strengthen this relationship. The contributions of this paper are as follows.
\begin{itemize}
\item We examined the relationship between gaze and task performance in the MR-DMT paradigm. Our experimental results reveal a weak gaze-performance link among children with ASD. 
\item We propose a novel dual-level visual guidance system that strengthens VMI by targeting perceptual and transformational processes, and validate its efficacy in significantly enhancing the gaze-performance link.
\end{itemize}

\section{Related Work}

\subsection{From Traditional to MR-DMT for children with ASD}

Dance Movement Therapy (DMT) is a psychotherapeutic intervention grounded in the principle of mind-body integration, aimed at promoting emotional, social, cognitive, and physical well-being~\cite{ADTA2009}. It is a holistic approach to healing, based on the empirically supported assertion that mind, body, and spirit are inseparable and interconnected; changes in one domain reflect and influence changes in another~\cite{ADTA2009}. DMT can promote various aspects of well-being in children with ASD~\cite{scharoun2014dance, aithal2021systematic}. Socially and emotionally, techniques such as mirroring help build rapport and enhance emotional understanding, potentially engaging mirror neuron mechanisms~\cite{mcgarry2011mirroring}. Physically, it improves motor coordination and spatial awareness~\cite{scharoun2014dance}. Crucially, it offers a non-verbal channel for expression, bypassing linguistic barriers often faced by individuals with ASD~\cite{scharoun2014dance}.

Despite its promise, traditional DMT faces practical constraints that limit its broader implementation. The approach is highly dependent on skilled therapists, resulting in challenges in scalability, accessibility, and cost-effectiveness. The limitations inherent in traditional DMT naturally motivate the exploration of innovative technological solutions. MR emerges as a particularly promising platform to directly address these challenges. Empirical studies have consistently shown that children with ASD exhibit significantly increased engagement and motivation in interventions utilizing immersive technologies like virtual reality (VR)~\cite{9419245, 10972798, 8446242}, augmented reality (AR)~\cite{premarathne2022dc}, and MR~\cite{liu2022evaluating, liu2021designing}. Unlike traditional clinic-based settings, MR systems can deploy virtual agents to deliver therapy, reducing reliance on scarce specialist therapists and lowering barriers to access. Liu et al. \cite{liu2024self} recently developed a groundbreaking self-guided MR-DMT system. Their system utilizes a virtual agent to guide children with ASD through the imitation of dance movements, and their results demonstrated significant improvements in performance, thereby providing robust initial evidence for the technical feasibility and efficacy of the MR-DMT paradigm.

\subsection{The Neurocognitive Barriers of Imitation-Based Intervention in ASD}

While MR-DMT~\cite{liu2024self} presents a promising intervention, its core task—imitation—directly targets the core deficits of ASD. ASD is clinically defined by persistent deficits in social communication and interaction, alongside restricted, repetitive patterns of behavior~\cite{edition1980diagnostic}. Beyond the core behavioral phenotype, general motor impairments are highly prevalent in ASD (90\%)~\cite{bhat2021motor}; moreover, motor imitation impairments~\cite{mostofsky2011altered, edwards2014meta, smith1994imitation} are consistently observed, and the degree of this impairment is predictive of core ASD symptoms~\cite{tunccgencc2021computerized, toth2006early}. The difficulty with imitation in ASD is hypothesized to stem from impairments in high-congruence VMI mechanisms~\cite{lidstone2021moving, williams2004systematic}. 

Deficits in VMI have been widely documented in children with ASD~\cite{chou2025visual, gepner2002rapid, miller2014dyspraxia, oliver2013visual, englund2014common}. VMI—defined as the ability to use visual information to guide motor actions~\cite{shin2009visuomotor}—involves the coordination of visual perception, visual processing, and fine motor skills~\cite{beery2004beery, carsone2023beery}. It supports preschool children in imitating movements, learning new skills, and engaging in social interactions. Research has also established a connection between VMI and both school readiness and adjustment \cite{decker2011cognitive, kulp1999relationship}. In children with ASD, VMI is significantly correlated not only with basic visual processes like discrimination, memory, and spatial relationships, but also with higher-order functions such as sequential memory and visual closure~\cite{chou2025visual}. 

VMI is commonly assessed through the ability to reproduce geometric forms, with the Beery Visual-Motor Integration (Beery VMI) test being the most widely used tool for this purpose~\cite{beery2004beery}. Children with ASD consistently demonstrate poorer performance on the Beery VMI~\cite{green2016beery, miller2014dyspraxia, rosenblum2019predictors} and other figure-copying tasks~\cite{englund2014common} compared to typically developing children. In addition, studies using computerized touch screen assessments further confirm that children with ASD exhibit difficulties in visuomotor integration~\cite{dowd2012planning}. This impairment suggests a change in neural circuitry that converts visual input into accurate motor responses.

The convergence of these deficits paints a clear picture of the neurocognitive barrier faced by imitation-based therapies such as DMT. And the therapeutic goal of DMT may conflict directly with the neurological profile of children with ASD. Therefore, any intervention, including an advanced MR-DMT system, that relies on imitation as its main learning mechanism must account for these VMI challenges. This critical context underscores the necessity of our study: to investigate whether these underlying neurocognitive challenges manifest as a weakened link between gaze and performance in an MR-DMT paradigm, and to explore how the intervention can be designed to compensate for and overcome these barriers.

\subsection{Gaze Guidance Strategies for ASD}

Given the barriers in VMI, merely presenting a demonstrative model in MR-DMT for imitation is insufficient. Therefore, interventions must actively guide the visual attention of children with ASD to effectively support VMI—a challenge we address in this section.

Due to the atypical gaze patterns~\cite{nakano2010atypical,drysdale2018gaze,ishizaki2021eye} commonly observed in children with ASD, numerous studies have investigated methods to direct and sustain their gaze with the goal of enhancing learning and visual attention. While early research on gaze guidance in immersive environments often focused on optimizing isolated metrics such as gaze duration~\cite{li2023use,wang2020promoting,liu2017feasibility}, recent work has begun to emphasize the functional impact of guided looking.

In a seminal study, Wang et al.~\cite{wang2020promoting} adopted a data-driven approach that utilized gaze patterns from typically developing (TD) children to construct normative attention models. Their system employed real-time eye tracking to dynamically modulate visual stimuli: when a child with ASD fixated on non-normative regions, those areas were perceptually degraded (e.g., via darkening or blurring), while areas aligned with TD attention patterns remained visually salient. This gaze-contingent modulation functioned as an implicit ``visual spotlight,'' effectively redirecting attention toward socially relevant features without explicit instruction.

Liu et al.~\cite{liu2017feasibility} demonstrated an AR-based strategy that used engaging cartoon overlays on real-world faces to attract and sustain visual attention. Their approach not only proved feasible but also led to measurable behavioral improvements, including enhanced eye contact, social engagement, and reductions in core ASD symptoms such as irritability and stereotypy.

Conversely, Li et al.~\cite{li2023use} implemented a VR-based attention training system incorporating real-time eye tracking and four operant conditioning strategies (positive/negative reinforcement and punishment) intended to increase focus on a virtual teacher. Surprisingly, their results showed that although gaze toward the teacher increased, overall distraction time also rose. This counterintuitive outcome underscores a critical concern: guided gaze behavior does not necessarily translate to functional performance gains. It emphasizes the need to design guidance systems that explicitly strengthen the gaze–performance link.

If prolonged gaze does not yield improved task performance, then gaze guidance alone is of limited value. Therefore, incorporating a gaze–performance link measure into intervention evaluations represents a potentially significant contribution to the field, shifting the focus from attentional processes alone to meaningful functional outcomes.

\section{Study 1: A Study Exploring Weak Gaze-Performance Link in MR-DMT}

Study 1 aims to explore the relationship between gaze and performance within imitation-based therapies. We test the question in MR-DMT paradigm for children with ASD, a virtual agent guides children through DMT sessions by demonstrating movements that the participants are instructed to imitate. 

\subsection{Hypothesis}

Based on the well-documented deficits in VMI among children with ASD, we hypothesize that in a baseline MR-DMT environment without visual guidance:

Hypothesis 1: The total gaze duration on task-relevant areas will show a weak and statistically non-significant correlation with overall motor performance.

We further hypothesize that this weak gaze-performance link is specific to the motor domain due to underlying VMI impairments. In contrast, for non-motor aspects of performance—such as engagement, affect, and task comprehension—gaze behavior is expected to exhibit a stronger association, as these domains rely more directly on visual attention rather than integrated motor output.

Hypothesis 2: Gaze duration will show a significantly stronger correlation with non-motor aspects of performance than with motor performance.

\subsection{Participants}
A total of 9 children with a clinical diagnosis of ASD (or ASD propensity) were recruited for study 1. Participants were aged between 2 and 6 years (M=4.56, SD=1.13), comprising 5 boys and 4 girls. Participants were recruited from a local hospital. All participating children exhibited normal hearing and vision capabilities. The study was conducted according to the guidelines of the Declaration of Helsinki and approved by the authors' institutional ethical review board (IRB) (No.XXX-XXX-XXXX-XXXX). Informed consent was obtained from the parents or guardians of each participant.

\subsection{Experimental Setup}

\begin{figure}[h]
 \centering 
 \vspace{-5pt} 
 \includegraphics[width=0.95\columnwidth]{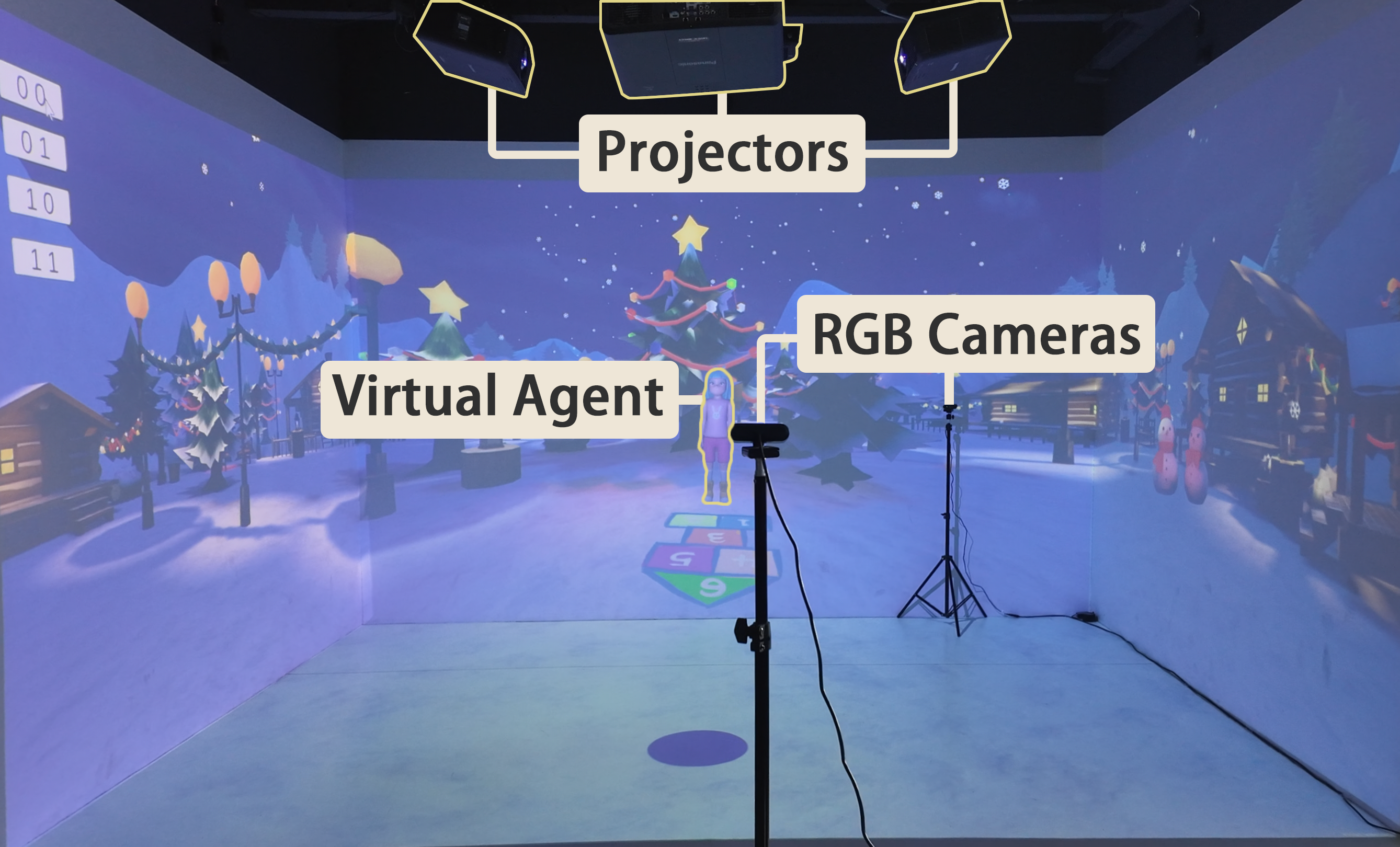}
 \caption{The baseline MR-DMT system in the CAVE environment.}
 \vspace{-10pt} 
 \label{fig:figure1}
\end{figure}

\label{Experimental Setup}
To investigate the relationship between gaze behavior and motor performance, a Mixed Reality Dance Movement Therapy (MR-DMT) system was implemented within a Cave Automatic Virtual Environment (CAVE) \cite{cruz1992cave} (see \autoref{fig:figure1}). This baseline system intentionally excluded all visual guidance strategies under evaluation in Study 2.

The CAVE structure measured 4.5 m × 3 m × 2.7 m and was equipped with four projectors and an integrated audio system to deliver immersive visual rendering across three walls and the floor. The MR-DMT application was developed in Unity (version 2021.3.12) and operated on a high-performance workstation (LEGION REN9000K) with an NVIDIA GeForce RTX 3070 Ti GPU, an Intel Core i7-12700KF processor, and 32 GB of RAM. Two RGB cameras were placed diagonally in front of and behind the child with ASD to capture multi-angle movement data for expert rating and behavioral coding.

A virtual agent, designed as a dance movement therapist taking the form of a virtual child, was displayed on the front CAVE wall to engage participants and provide step-by-step demonstrations of dance motions. Using a decomposed teaching method, the agent broke down each movement into manageable segments to facilitate progressive motor learning. The children were instructed to imitate the agent’s movements. 

\subsection{Measures}
\label{Measures}

To test the hypotheses regarding the differential relationship between gaze behavior and motor versus non-motor performance, the following variables were quantitatively assessed: (1) total gaze duration on task-relevant areas, (2) motor imitation performance, and (3) target-related responsive behaviors. Expert ratings were provided by a panel of three therapists specializing in ASD rehabilitation from a local hospital, with clinical experience ranging from 9 to 16 years (M = 11.67, SD = 3.79) and ages between 32 and 38 years (M = 35.67, SD = 10.33).

\subsubsection{Total Gaze Duration on Task-Relevant Areas}

Gaze behavior was operationalized as the total duration of visual fixation on task-relevant areas (e.g., the virtual agent’s body parts or instructional aids). The expert panel performed behavioral coding based on recorded session videos. The coders identified and annotated the start and end timestamps of each fixation event, from which the total and average gaze duration per participant were derived.

\subsubsection{Performance}

We evaluated participant performance across both motor and non-motor domains through expert ratings.

\textbf{Motor Imitation Performance:}Consistent with H1, motor performance was evaluated to examine its link with gaze behavior. Following the framework in \cite{liu2024self}, the experts assessed four dimensions of movement execution: accuracy, rhythm, coordination, and completeness. Each dimension was rated on an 11-point scale (0–10) through repeated viewing of session recordings. High inter-rater reliability was achieved across all motor performance ratings.

\textbf{Target-Related Responsive Behaviors:}To evaluate non-motor aspects of performance relevant to H2, we administered a post-training questionnaire adapted from \cite{liu2021designing}. This instrument measured eight behavioral and affective dimensions: (1) ability to participate independently; (2) concentration during training; (3) flexibility in movement and operation; (4) happiness after task completion; (5) performance in listening and repeating; (6) degree of movement reproduction; (7) understanding of course content; and (8) expression of positive emotions. Each item was rated by the expert panel on a 5-point Likert scale (1 = strongly disagree to 5 = strongly agree). The questionnaire has demonstrated good reliability in previous studies.

\subsection{Procedure}

\begin{figure}[!h]
 \centering 
 \vspace{-10pt} 
 \includegraphics[width=1\columnwidth]{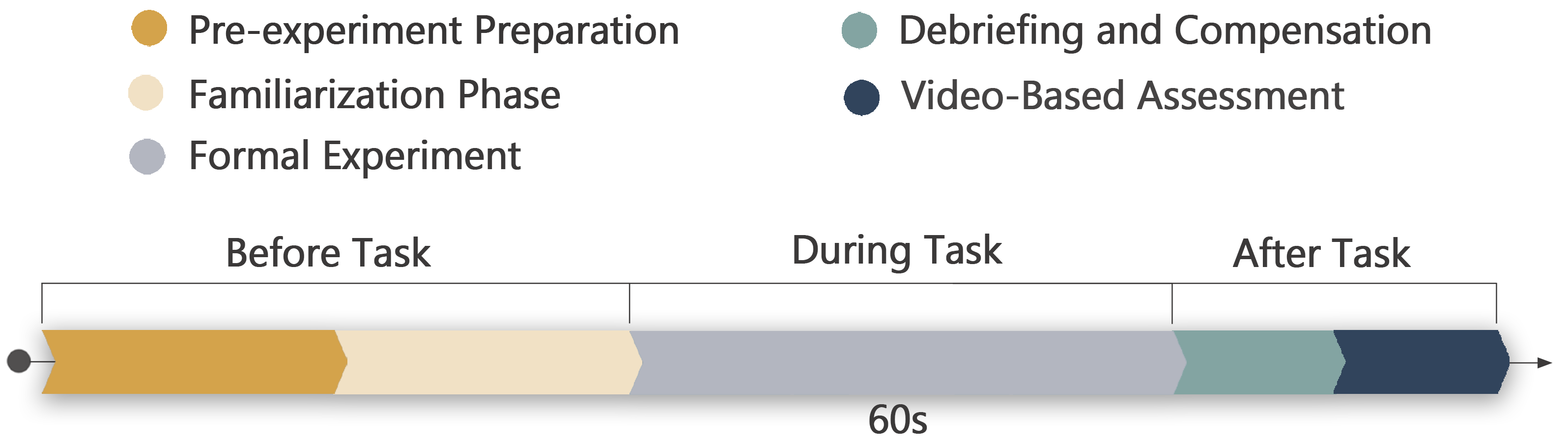}
 \caption{The experimental procedure of study 1.}
 \vspace{-10pt} 
 \label{fig:figure2}
\end{figure}

The MR-DMT system was configured as described in Section \ref{Experimental Setup}. The overall procedure is summarized in Figure \ref{fig:figure2}. The steps are detailed below:

\begin{enumerate} 
\item Pre-experiment preparation: Parents or guardians provided informed consent and completed a demographic questionnaire.
\item Familiarization phase: Participants engaged in a practice session with basic movements to acclimatize to the MR environment before the formal task. The practice session lasted approximately two and a half minutes.
\item Formal experiment: Each participant took part in the MR-DMT session, during which a virtual agent demonstrated and guided the participant through a complete dance routine following a children’s song. Parents were seated slightly behind the participant to monitor safety without interfering. Each session lasted approximately one minute.
\item Post-experiment: Upon completion, each participant received a small gift as a token of appreciation.
\item Video analysis: Expert rating and behavioral coding were performed offline by the assessment panel based on the recorded session videos, following the methodology detailed in Section \ref{Measures}.
\end{enumerate}

\subsection{Results}

\begin{table}[htb]
\vspace{-2pt}
\centering
\caption{Mean and intercorrelations for variables in Study 1.}
\label{tab:gaze_correlations}
\footnotesize
\setlength{\tabcolsep}{3pt} 
\renewcommand{\arraystretch}{1.2} 
\begin{tabular}{@{} L{5.0cm} C{1cm} C{1.1cm} @{}}
\toprule
\textbf{Variable} & \textbf{Mean (SD)} & \textbf{$\textrho$} \\
\midrule
\textbf{Total gaze duration on task-relevant areas} & \textbf{17.37 (11.60)} & \textbf{1 (0.000***)} \\
\midrule
\textbf{Motor imitation performance} & \textbf{5.41 (5.85)} & \textbf{0.536 (0.137)} \\
Accuracy & 1.48 (1.72) & 0.521 (0.150) \\
Rhythm & 1.11 (1.27) & 0.635 (0.066) \\
Coordination & 1.41 (1.53) & 0.554 (0.122) \\
Completeness & 1.41 (1.55) & 0.536 (0.137) \\
\midrule
\textbf{Target-related responsive behaviors} & \textbf{14.78 (6.65)} & \textbf{0.525 (0.147)} \\
Ability to participate independently & 2.48 (1.73) & 0.536 (0.137) \\
Concentration during training & 2.37 (1.26) & 0.661 (0.053) \\
Flexibility in movement and operation & 1.44 (0.73) & 0.84 (0.005**) \\
Happiness after task completion & 1.04 (0.11) & -0.413 (0.270) \\
Performance in listening and repeating & 1.33 (0.44) & 0.636 (0.066) \\
Degree of movement reproduction & 1.59 (0.78) & 0.563 (0.114) \\
Understanding of course content & 1.56 (0.99) & 0.697 (0.037*) \\
Expression of positive emotions during the process & 2.78 (1.54) & 0.581 (0.101) \\
\bottomrule
\end{tabular}
\vspace{0.2cm}

\footnotesize\textit{ Note: *** : $p < .001; ** : p < .01; * : p < .05; M = Mean, SD = Standard~Deviation
$. $\textrho~$denotes Spearman's correlation coefficient between total gaze duration on task-relevant areas and each variable.}
\end{table}

\begin{figure}[!h]
 \centering 
 \includegraphics[width=0.95\columnwidth]{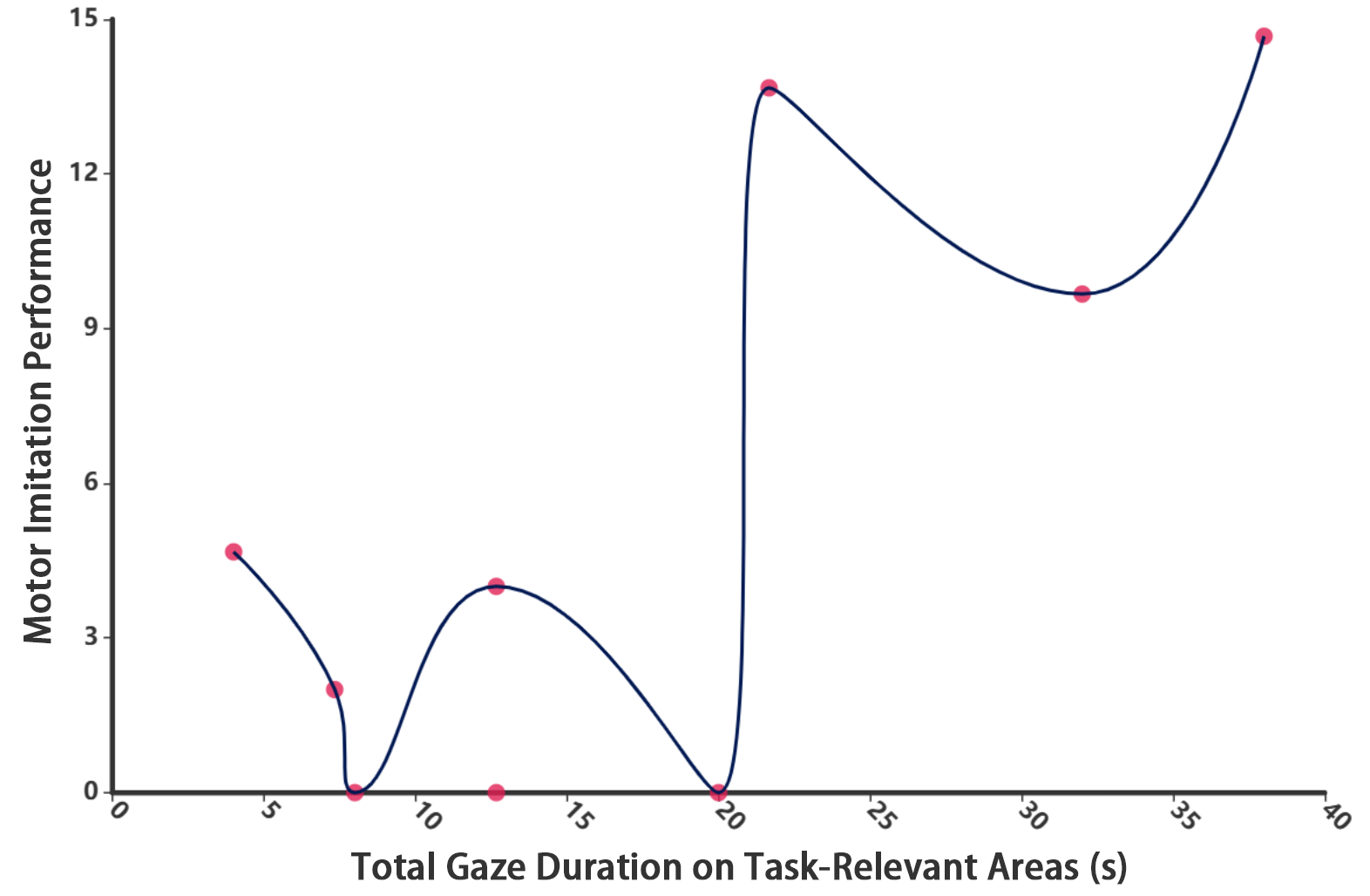}
 \caption{Scatter plot of total gaze duration on task-relevant areas and motor imitation performance scores in the baseline condition of Study 1.}
 \vspace{-10pt} 
 \label{fig:figure8}
\end{figure}

Descriptive statistics and Spearman’s rank correlation analyses were conducted for all study variables. Table \ref{tab:gaze_correlations} presents the means, standard deviations (SD), and intercorrelations among the variables.

As shown in the table, no significant correlations were found between total gaze duration on task-relevant areas and any of the four indicators of motor imitation performance, nor with the majority of target-related responsive behaviors—supporting Hypothesis 1. However, two exceptions were observed among non-motor outcomes: a strong positive correlation was identified between total gaze duration and ``flexibility in movement and operation'' ($\textrho = 0.84$, $p < .01$), as well as a moderate positive correlation with ``understanding of course content'' ($\textrho = 0.697$, $p < .05$).

The weak gaze-performance link observed in Study 1, as visually evidenced by the scattered plot of individual data points (Figure \ref{fig:figure8}). It is important to note that this finding is based on a small sample size; however, the successful replication of this effect in the baseline condition of Study 2 (with a larger cohort) strengthens our confidence in its reliability.

Overall, these results provide general support for both Hypothesis 1 and Hypothesis 2, indicating a dissociation between gaze duration and motor performance, while suggesting a more nuanced relationship with certain non-motor, responsive behaviors.

\subsection{Discussion}

The primary objective of Study 1 was to investigate the fundamental relationship between gaze behavior and task performance within a non-guided MR-DMT environment for children with ASD. As hypothesized, our findings provide strong empirical support for the hypothesized \textit{weak gaze-performance link} in the motor domain, while also revealing a more nuanced pattern in non-motor outcomes.

Consistent with H1, we found no significant correlation between total gaze duration on task-relevant areas and motor imitation performance. This result aligns with the well-documented VMI deficits characteristic of ASD, whereby the extended duration of gaze does not readily translate into improved motor performance. Despite sufficient visual attention to the virtual agent’s demonstrations, participants struggled to convert this perceptual input into accurate motor output—a dissociation that underscores the core neurocognitive barrier limiting imitation-based interventions for this population.

In partial support of H2, gaze duration showed a stronger association with certain non-motor, responsive behaviors than with motor performance. Specifically, significant correlations were found with ``flexibility in movement and operation'' and ``understanding of course content'', suggesting that visual attention may contribute more directly to cognitive and adaptive aspects of performance than to motoric ones. This differentiation reinforces the notion that not all performance dimensions rely equally on integrated visuomotor processing. Whereas motor imitation heavily depends on the integrity of VMI pathways, comprehension and flexibility may rely more heavily on visual-spatial and attentional resources, which are less impaired in some children with ASD.

Notably, the majority of target-related responsive behaviors—such as happiness after task completion, ``expression of positive emotions,'' and ``ability to participate independently''—were not significantly correlated with gaze. This indicates that even within non-motor domains, gaze behavior alone is not a universal predictor of therapeutic engagement or affective response. These outcomes may be influenced by factors beyond visual attention, including intrinsic motivation, sensory preferences, and emotional regulation.

Together, these findings validate our central argument: that \textit{simply increasing look time is insufficient to improve performance in MR-DMT}, and that interventions must explicitly target the gaze-performance link rather than gaze behavior alone. The identified dissociation highlights the critical need for guidance strategies that not only attract attention but also facilitate its conversion into motor execution—a challenge we aim to address in Study 2.

\section{Study 2: A STUDY OF A POTENTIAL APPROACH TO STRENGTHEN THE WEAK LINK}

Building on the weak gaze-performance link identified in Study 1, this study proposed and evaluated a novel dual-level visual guidance system designed to strengthen the VMI process in children with ASD during MR-DMT. We hypothesized that this guidance paradigm would not only increase gaze toward task-relevant areas but, more importantly, would significantly enhance the correlation between gaze duration and motor imitation performance.

\subsection{Design Rationale and Objectives}

Study 1 confirmed that merely attracting gaze is insufficient to improve motor performance in children with ASD, highlighting a critical deficit in VMI. Therefore, the objective of Study 2 was to design an intervention that actively facilitates the conversion of visual attention into motor action.

Our proposed solution is a dual-level visual guidance system implemented within the same MR-DMT paradigm. Unlike prior approaches that optimize gaze in isolation, our system is theorized to operate at two levels:
\begin{enumerate}
\item Perceptual Level: To direct attention to the correct spatial locations (the where).
\item Transformational Level: To scaffold the cognitive-motor translation from observation to execution (the how).
\end{enumerate}
We hypothesized that:

Hypothesis 3: Participants in the dual-level visual guidance condition will show a significantly stronger correlation between total gaze duration on task-relevant areas and performance than those in a baseline (no-guidance) condition.

\subsection{The Dual-Level Visual Guidance System}

\begin{figure}[!th]
 \centering 
 \includegraphics[width=1\columnwidth]{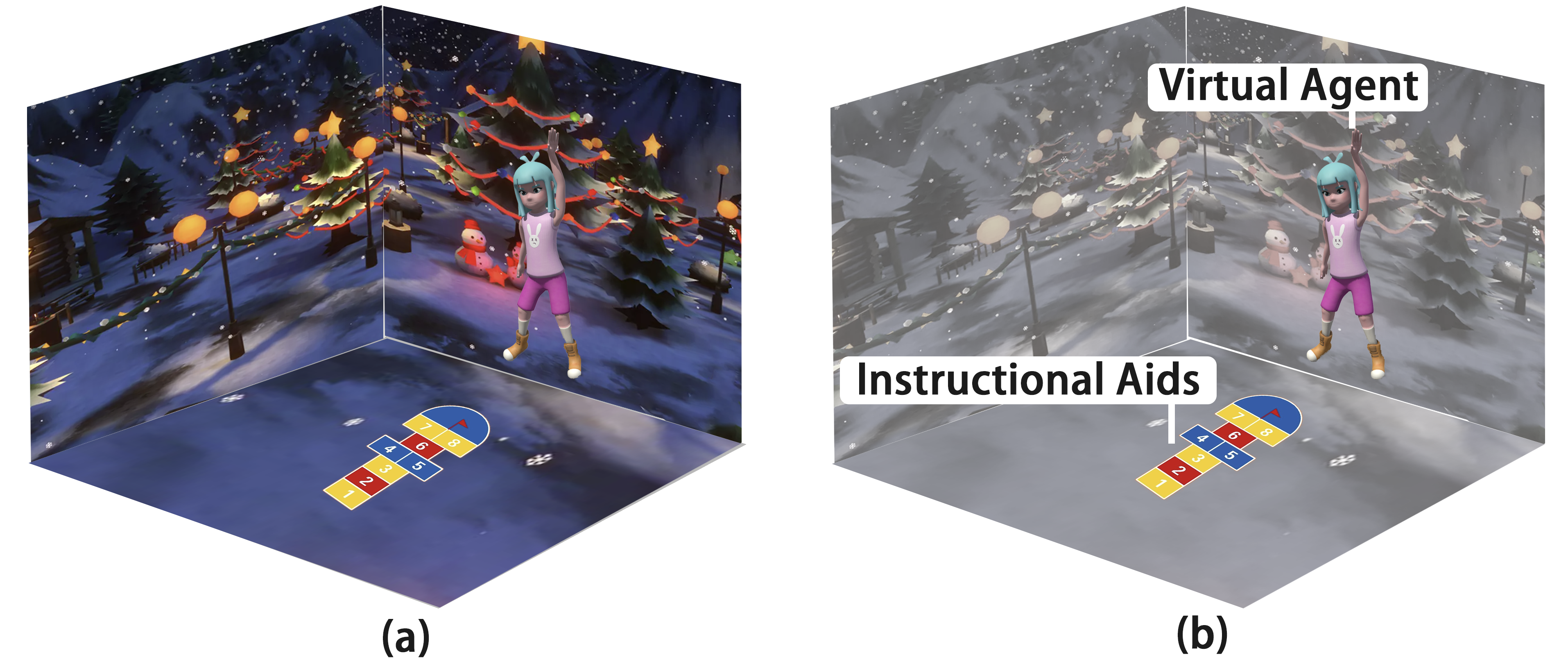}
 \caption{Visual comparison of the perceptual guidance manipulation. (a) Perceptual Guidance absent: All elements in the environment retain their original visual properties without modulation. (b) Perceptual Guidance present: The saturation and brightness of task-irrelevant background regions are reduced, creating a spotlight effect that highlights the virtual agent and instructional aids.}
 \label{fig:figure6}
\end{figure}

\begin{figure}[!th]
 \centering 
 \includegraphics[width=1\columnwidth]{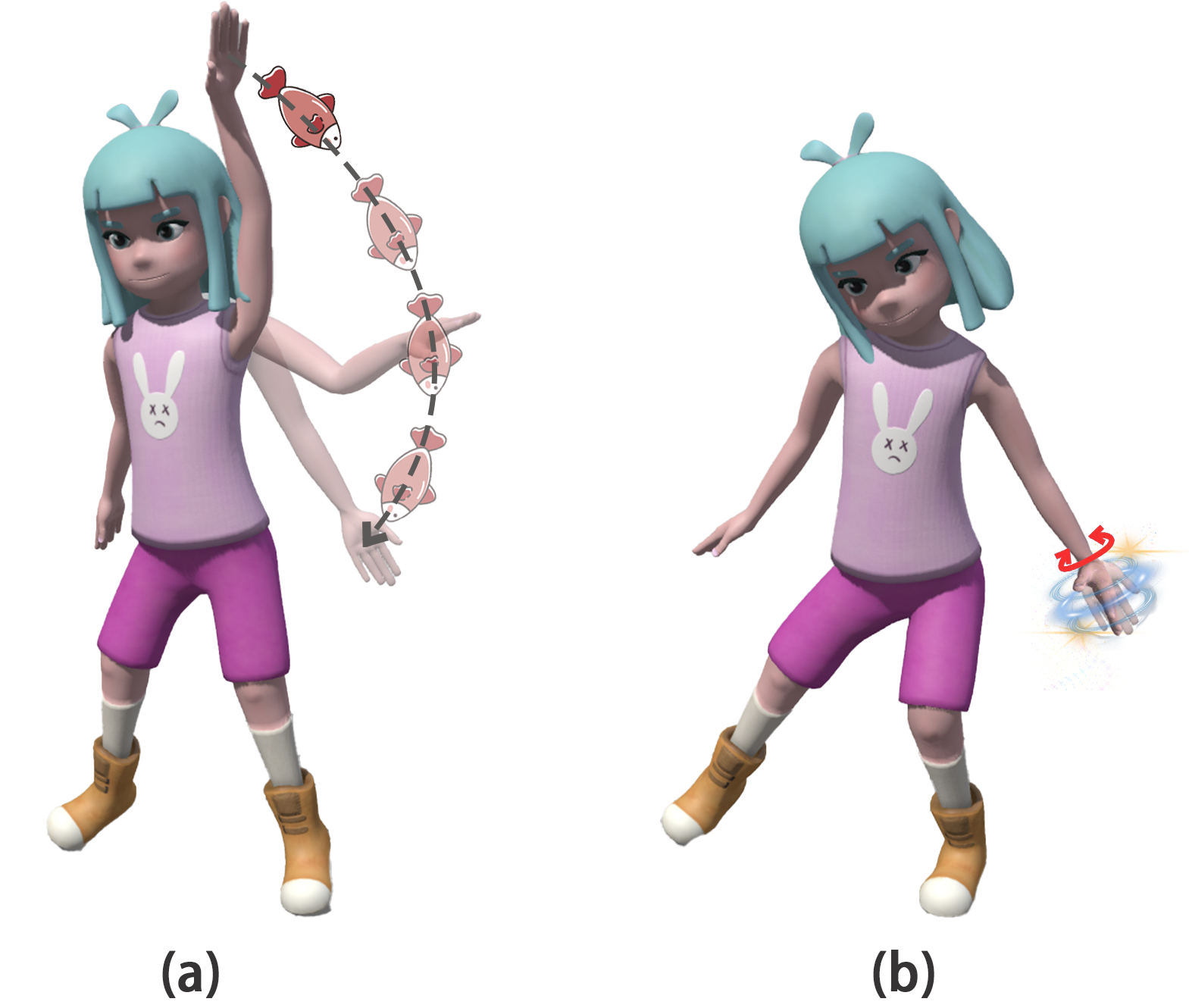}
 \caption{Examples of transformational guidance through kinetic metaphors. (a) A school of fish animates along the trajectory of the virtual agent's arm movement to provide a predictive cue. (b) Dynamic glow effects are applied to emphasize the goal of a complex dance motion.}
 \vspace{-20pt} 
 \label{fig:figure7}
\end{figure}

To address the weak gaze-performance link identified in Study 1, we designed a dual-level visual guidance system that targets both perceptual engagement and motor transformation stages of visuomotor integration (VMI). This system operates through two complementary mechanisms:

\textbf{Level 1: Perceptual Guidance through Salience Modulation.}
This level directs visual attention explicitly toward task-relevant elements. Based on the principle of perceptual salience, we reduced the saturation and brightness of all task-irrelevant regions in the CAVE environment. Meanwhile, the virtual agent and instructional aids retained their original visual properties, keeping them clearly visible. This approach created a subdued background, making the key teaching elements stand out visually. This formed a gentle ``spotlight'' that guided attention smoothly, without causing disruption or cognitive overload. The visual contrast between conditions with and without this perceptual guidance is illustrated in Figure \ref{fig:figure6}, where (a) shows the baseline condition without such modulation, and (b) demonstrates the effect of salience modulation with irrelevant regions dimmed.

\textbf{Level 2: Transformational Guidance through Kinetic Metaphors.}
To scaffold the translation of observed movement into motor execution, this level provides explicit kinematic cues using metaphor-based animation. Specifically, we animated a school of fish that swam along the trajectory of the virtual agent’s limb movements—for example, when the agent raised its arm, the fish would swim upward along the same path (see \autoref{fig:figure7} (a)). Prior studies found that children with ASD show a strong interest in fish within MR environments. We adapted this insight by using animated fish to provide movement guidance. These kinetic metaphors serve as predictive visual cues that help children anticipate movement direction, understand movement goals, and plan their own motor responses. Additionally, for particularly complex or unusual movements, supplementary effects such as dynamic glow or subtle flashing were applied in the direction of the motion to further emphasize movement intent and enhance comprehensibility (see \autoref{fig:figure7} (b)).

Together, these two levels of guidance form an integrated system that not only directs visual attention to the correct spatial locations (Level 1) but also disambiguates movement meaning and supports motor planning (Level 2), thereby directly facilitating the VMI process.

\subsection{Experimental Design}
We employed a 2 × 2 within-subjects design to evaluate the effects of our dual-level visual guidance system. The two factors were:
\begin{itemize}
\item Perceptual Guidance (two levels: present vs. absent)
\item Transformational Guidance (two levels: present vs. absent)
\end{itemize}
Each participant experienced all four experimental conditions in a counterbalanced order. The dependent variable was Spearman’s correlation coefficient between the gaze duration and other variables across conditions.

In total, there were four experimental conditions:
\begin{itemize}
\item \textbf{Condition 1: Baseline} - Both perceptual and transformational guidance \textbf{absent}. This condition used the same MR-DMT system configuration as in Study 1, but featured a different dance routine and song.
\item \textbf{Condition 2: Perceptual Only} - Perceptual guidance \textbf{present} , transformational guidance \textbf{absent}.
\item \textbf{Condition 3: Transformational Only} - Transformational guidance \textbf{present}, perceptual guidance \textbf{absent}.
\item \textbf{Condition 4: Dual-level Guidance} - Both perceptual and transformational guidance \textbf{present}, representing the complete multi-level guidance system.
\end{itemize}
Each participant experienced all four conditions. We employed a Latin Square design to counterbalance the potential order effects of experiencing the four conditions.

\subsection{Participants}

A total of 48 children with a clinical diagnosis of ASD (or ASD propensity) were recruited for Study 2; none had participated in Study 1. From this cohort, a total of 18 valid data points were collected for analysis. The final sample size was reduced from the initial recruitment total primarily due to factors common in ASD research with young children, including sensory overload from the equipment, and inability to complete all experimental conditions. Participants ranged in age from 2 to 7 years (mean = 11.06, SD = 2.61), including 10 boys and 8 girls. All children were recruited through a collaborating local hospital and had normal hearing and vision capabilities, as reported by parents or guardians. The study was conducted in accordance with the ethical principles outlined in the Declaration of Helsinki and was approved by the Institutional Review Board of the authors (IRB Approval No.: XXX-XXX-XXXX-XXXX). Written informed consent was obtained from the parent or legal guardian of each participant before the study.

\begin{table*}[th]
\vspace{-0.5cm}
\centering
\caption{Means, standard deviations, and Spearman's correlations between gaze duration and outcome measures across guidance conditions.}
\label{tab:guidance_comparison}
\footnotesize
\setlength{\tabcolsep}{4pt}
\renewcommand\arraystretch{1.1}

\begin{tabular}{@{} L{2.8cm} *{8}{C{1.3cm}} @{}}
\toprule
\multirow{2}{*}{Variables} & 
\multicolumn{2}{c}{\makecell{Baseline}} & 
\multicolumn{2}{c}{\makecell{Perceptual Only}} & 
\multicolumn{2}{c}{\makecell{Transformational\\Only}} & 
\multicolumn{2}{c}{\makecell{Dual-level\\Guidance}} \\
\cmidrule(lr){2-3} \cmidrule(lr){4-5} \cmidrule(lr){6-7} \cmidrule(lr){8-9}
& M(SD) & r & M(SD) & r & M(SD) & r & M(SD) & r \\
\midrule

\textbf{Total gaze duration} & \textbf{68.46} & \textbf{1} & \textbf{83.07} & \textbf{1}  & \textbf{83.57} & \textbf{1} & \textbf{81.44} & \textbf{1} \\
\textbf{on task-relevant areas} & \textbf{(23.86)} &  \textbf{(0.000***)}  & \textbf{(30.06)} & \textbf{(0.000***)} & \textbf{(31.14)} & \textbf{(0.000***)} & \textbf{(31.86)} &  \textbf{(0.000***)}\\
\midrule

\textbf{Motor imitation} & \textbf{20.67} & \textbf{0.392} & \textbf{19.93} & \textbf{0.581} & \textbf{22.89} & \textbf{0.556} & \textbf{22.69} & \textbf{0.731} \\
\textbf{performance} & \textbf{(9.43)} & \textbf{(0.107)} & \textbf{(7.50)} & \textbf{(0.011*)} & \textbf{(8.58)} & \textbf{(0.017*)} & \textbf{(9.43)} & \textbf{(0.001***)} \\\\[0.1em]

Accuracy & 5.37 & 0.491 & 5.87 & 0.543 & 5.89 & 0.547 & 5.87 & 0.794 \\
 & (2.43) & (0.039*) & (2.32) & (0.02*) & (2.18) & (0.019*) & (2.42) & (0.000***) \\

Rhythm & 4.75 & 0.333 & 5.37 & 0.539 & 5.44 & 0.513 & 5.41 & 0.634 \\
 & (2.21) & (0.176) & (2.19) & (0.021*) & (1.98) & (0.029*) & (2.29) & (0.005**) \\

Coordination & 5.02 & 0.331 & 5.46 & 0.457 & 5.48 & 0.537 & 5.48 & 0.761 \\
 & (2.28) & (0.18) & (2.28) & (0.057) & (2.11) & (0.021*) & (2.26) & (0.000***) \\

Completeness & 5.54 & 0.437 & 5.93 & 0.522 & 6.07 & 0.488 & 5.93 & 0.702 \\
 & (2.63) & (0.07) & (2.43) & (0.026*) & (2.41) & (0.04*) & (2.56) & (0.001***) \\
\midrule
\textbf{Target-related} & \textbf{25.49} & \textbf{0.486} & \textbf{27.09} & \textbf{0.498} & \textbf{27.26} & \textbf{0.611} & \textbf{27.02} & \textbf{0.743} \\
\textbf{responsive behaviors} & \textbf{(7.68)} & \textbf{(0.041*)} & \textbf{(5.97)} & \textbf{(0.035*)} & \textbf{(5.40)} & \textbf{(0.007**)} & \textbf{(5.98)} & \textbf{(0.000***)} \\\\[0.1em]

Ability to participate & 3.71 & 0.707 & 3.91 & 0.563 & 3.98 & 0.586 & 3.93 & 0.722 \\
independently & (1.04) & (0.001***) & (0.71) & (0.015*) & (0.66) & (0.011*) & (0.65) & (0.001***) \\\\[0.1em]

Concentration during & 3.15 & 0.692 & 3.31 & 0.613 & 3.39 & 0.617 & 3.46 & 0.713 \\
training & (0.96) & (0.001***) & (0.75) & (0.007**) & (0.83) & (0.006**) & (0.77) & (0.001***) \\\\[0.1em]

Flexibility in & 3.07 & 0.495 & 3.28 & 0.401 & 3.28 & 0.471 & 3.24 & 0.78 \\
movement and operation & (0.95) & (0.037*) & (0.81) & (0.099) & (0.74) & (0.048*) & (0.91) & (0.000***) \\\\[0.1em]

Happiness after & 3.37 & 0.304 & 3.67 & 0.439 & 3.59 & 0.504 & 3.52 & 0.606 \\
task completion & (0.98) & (0.22) & (0.81) & (0.068) & (0.74) & (0.033*) & (0.82) & (0.008**) \\\\[0.1em]

Performance in & 2.75 & 0.48 & 2.87 & 0.586 & 2.87 & 0.513 & 2.81 & 0.794 \\
listening and repeating & (1.00) & (0.044*) & (0.84) & (0.011*) & (0.81) & (0.029*) & (0.81) & (0.000***) \\\\[0.1em]

Degree of movement & 2.86 & 0.369 & 3.09 & 0.51 & 3.09 & 0.468 & 3.17 & 0.662 \\
reproduction & (1.03) & (0.131) & (0.89) & (0.031*) & (0.85) & (0.05) & (1.04) & (0.003**) \\\\[0.1em]

Understanding of & 3.00 & 0.422 & 3.20 & 0.349 & 3.15 & 0.53 & 3.19 & 0.766 \\
course content & (1.06) & (0.081) & (0.89) & (0.155) & (0.81) & (0.024*) & (0.83) & (0.000***) \\\\[0.1em]

Expression of positive & 3.61 & 0.448 & 3.76 & 0.475 & 3.91 & 0.636 & 3.70 & 0.44 \\
emotions during process & (0.95) & (0.063) & (0.62) & (0.046*) & (0.41) & (0.005**) & (0.61) & (0.068) \\

\bottomrule
\end{tabular}

\footnotesize\textit{ Note: *** : $p < .001; ** : p < .01; * : p < .05; M = Mean, SD = Standard~Deviation
$. $\textrho~$denotes Spearman's correlation coefficient between total gaze duration on task-relevant areas and each variable.}
\vspace{-0.5cm}
\end{table*}

\subsection{Measures}

The same measures as those used in Study 1 were collected to directly assess the impact of over guidance on the gaze-performance link. These included: (1)Total Gaze Duration on Task-Relevant Areas, (2)Motor Imitation Performance, and (3)Target-Related Responsive Behaviors.

\subsection{Procedure}

\begin{figure}[!th]
 \centering 
 \vspace{-10pt} 
 \includegraphics[width=1\columnwidth]{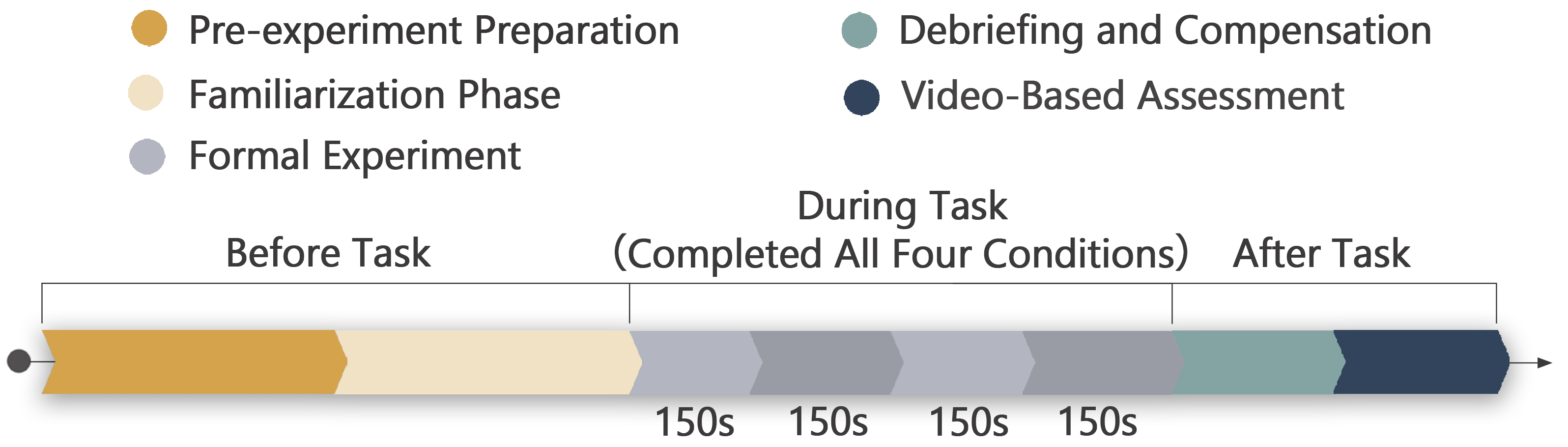}
 \caption{The experimental procedure of study 2.}
 \vspace{-10pt} 
 \label{fig:figure4}
\end{figure}

\begin{figure}[!th]
 \centering 
 \vspace{-2pt} 
 \includegraphics[width=1\columnwidth]{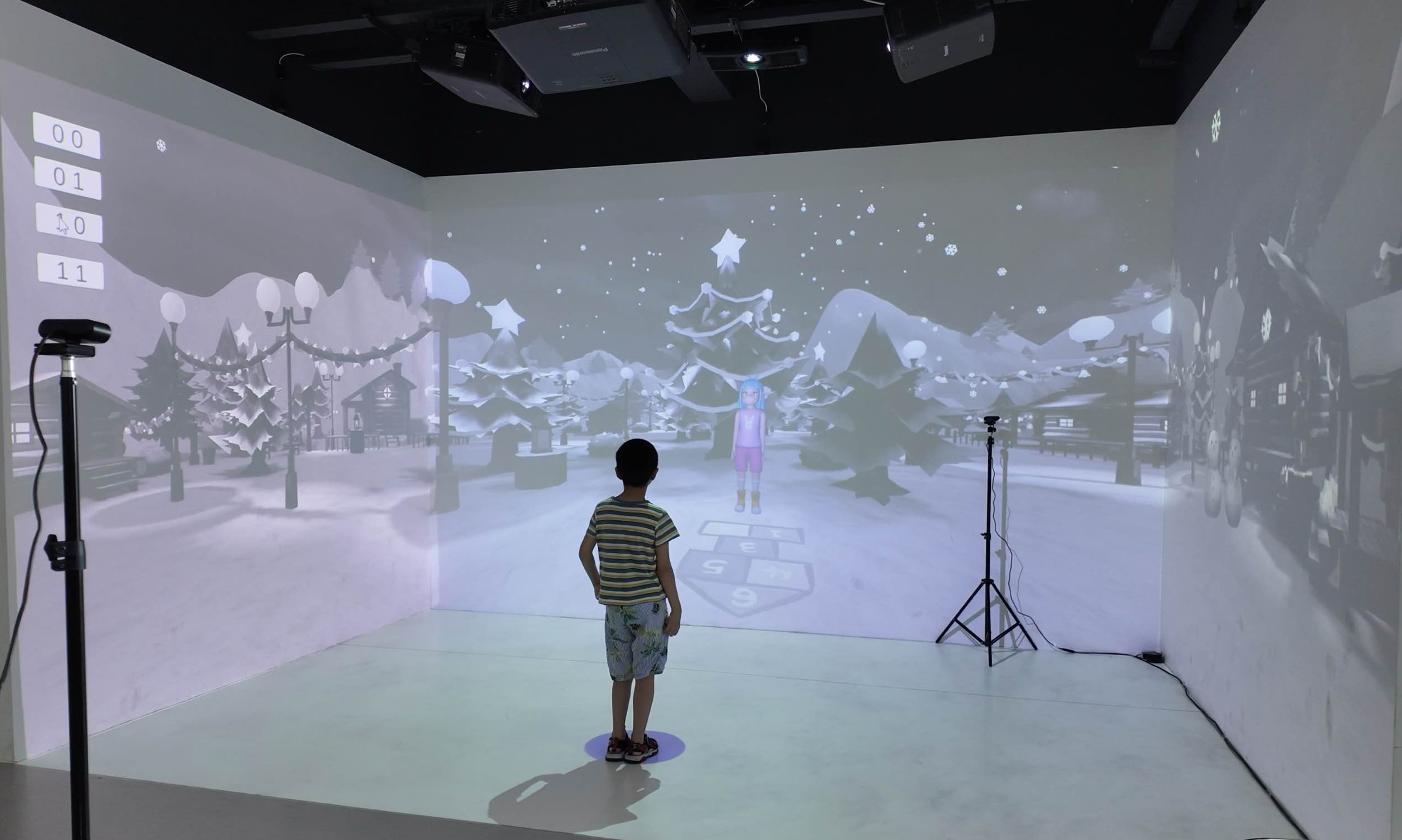}
 \caption{A snapshot of the experimental scenario.}
 \vspace{-10pt} 
 \label{fig:figure5}
\end{figure}

The MR-DMT system was configured according to the experimental design prior to the study. A schematic overview of the experimental procedure is provided in Figure \ref{fig:figure4}. The scene of the experimental site is shown in Figure \ref{fig:figure5}. The specific steps are detailed below:

\begin{enumerate}
\item Pre-experiment preparation: Before the experiment, parents or guardians signed the informed consent form and completed a demographic questionnaire.
\item Familiarization phase: Participants participated in a practice session consisting of basic movements to acclimate to the MR-DMT environment. This session lasted approximately 2.5 minutes.
\item Formal experiment: Each participant subsequently completed the MR-DMT session under each of the four experimental conditions, with order counterbalanced across participants. A parent or guardian remained seated slightly behind the participant throughout to ensure safety. The total duration of the experimental tasks was approximately 10 minutes.
\item Post-experiment: Upon completion, each participant received a small gift as a token of appreciation.
\item Video analysis: Expert rating and behavioral coding were performed offline based on the recorded session videos.
\end{enumerate}

\subsection{Results}

\begin{figure*}[!th]
 \centering \centering instead (more compact)
 \includegraphics[width=\textwidth]{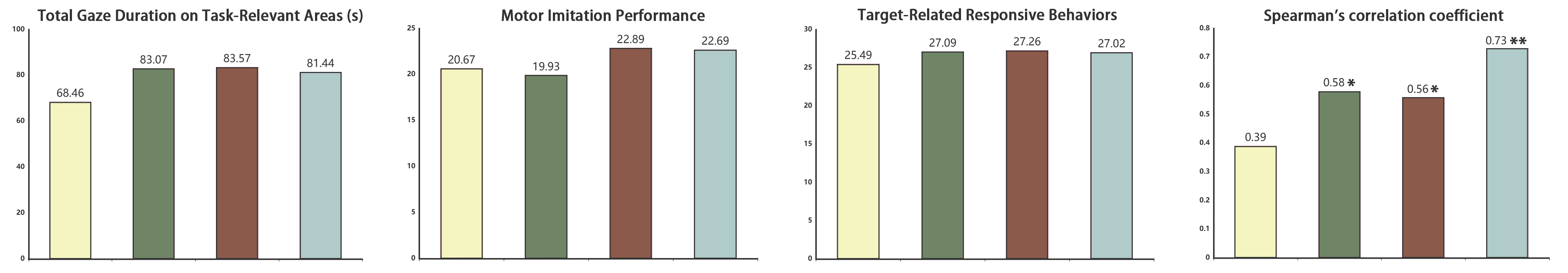}
 \caption{Comparison of key metrics across guidance conditions: total gaze duration, motor imitation performance, target-related responsive behaviors, and their correlation strength.}
 \vspace{-15pt} 
 \label{fig:figurez}
\end{figure*}

Descriptive statistics and Spearman’s rank correlation analyses were conducted for all study variables. Table \ref{tab:guidance_comparison} presents the means, standard deviations (SD), and intercorrelations among the variables.

The differential effects of the four guidance strategies are visually summarized in \autoref{fig:figurez}, which presents a comprehensive comparison of the primary outcome measures: (1) total gaze duration on task-relevant areas, (2) motor imitation performance scores, (3) target-related responsive behavior scores, and (4) the strength of the gaze-performance link (Spearman's $\rho$). Visual inspection of the figure reveals that the Dual-level Guidance condition consistently produced the most favorable outcomes across multiple metrics.

\subsubsection{Effects on Gaze Duration and Motor Performance}

Total gaze duration on task-relevant areas was highest in the \textit{With Global-Without Local} condition ($M = 83.57$, $SD = 31.14$), though all guidance conditions resulted in longer gaze durations compared to the baseline ($M = 68.46$, $SD = 23.86$). Most notably, the correlation between gaze duration and motor imitation performance strengthened progressively across conditions. In the \textit{Baseline} condition (no guidance), the correlation was non-significant ($\rho = 0.392$, $p = 0.107$). The introduction of either guidance type significantly strengthened this association: \textit{Without Global-With Local} ($\rho = 0.581$, $p = 0.011$), \textit{With Global-Without Local} ($\rho = 0.556$, $p = 0.017$). The strongest correlation was observed in the dual-level guidance condition \textit{ with global-with local} ($\rho = 0.731$, $p < 0.001$), indicating a substantial enhancement in the gaze performance link.

This strengthening pattern was consistent across all four motor dimensions (accuracy, rhythm, coordination, completeness), with the strongest correlations again emerging in the dual-level guidance condition (e.g., accuracy: $\rho = 0.794$; coordination: $\rho = 0.761$; both $p < 0.001$).

In addition to the strengthening of the gaze-performance correlation, we observed a numerical improvement in mean motor imitation scores from the Baseline ($M = 20.67$, $SD = 9.43$) to the Dual-level Guidance condition ($M = 22.69$, $SD = 9.43$). The magnitude of this difference corresponds to a Cohen's $d = 0.21$, indicating a small effect.

\subsubsection{Effects on Target-Related Responsive Behaviors}

A similar strengthening effect was observed in the relationship between gaze duration and target-related responsive behaviors. The correlation was significant but moderate in the baseline condition ($\rho = 0.486$, $p = 0.041$) and increased in strength with the addition of guidance, reaching its peak in the dual-level guidance condition ($\rho = 0.743$, $p < 0.001$). Analysis of subscales revealed that abilities such as \textit{Ability to participate independently} and \textit{Concentration during training} showed robust correlations across all conditions (e.g., ranging from $\rho = 0.563$ to $0.722$), while affective measures like \textit{Happiness after task completion} and \textit{Expression of positive emotions} exhibited more variable and generally weaker associations.

Similarly, for target-related responsive behaviors, the mean score was higher under Dual-level Guidance ($M = 27.02$, $SD = 5.98$) than at Baseline ($M = 25.49$, $SD = 7.68$), with a Cohen's $d$ of $0.22$, also indicating a small effect.

\subsubsection{Replication of Study 1 Findings in Baseline Condition}

Critically, the data from the \textit{Baseline} condition (no guidance) in Study 2 successfully replicated the core findings of Study 1, providing further validation for its hypotheses. Mirroring the results of Study 1, the correlation between total gaze duration and overall motor imitation performance in the baseline condition was weak and statistically non-significant ($\rho = 0.392$, $p = 0.107$), providing strong convergent support for \textbf{Hypothesis 1}.

Furthermore, the pattern of correlations in the baseline condition also corroborated \textbf{Hypothesis 2}. Consistent with Hypothesis 2, the gaze-performance link was weak in the motor domain, whereas gaze duration exhibited significantly stronger correlations with several non-motor, responsive behaviors. The overall correlation with the composite score of target-related responsive behaviors was $\rho = 0.486$ ($p = 0.041$), and notably stronger correlations were observed with specific aspects such as \textit{Ability to participate independently} ($\rho = 0.707$, $p < 0.001$) and \textit{Concentration during training} ($\rho = 0.692$, $p < 0.001$). This dissociation—weak association with motor performance but stronger associations with specific non-motor, target-related responsive behaviors—provides additional confirmation of the hypothesized differential relationship between gaze behavior and various performance domains.

This successful replication within the baseline condition of Study 2 not only strengthens the validity of the initial findings but also firmly establishes the weak gaze-performance link as a reliable phenomenon in non-guided MR-DMT settings for children with ASD, thereby underscoring the necessity for targeted interventions.

\subsubsection{Summary of Key Findings}

The findings from Study 2 provide strong empirical support for the effectiveness of our dual-level visual guidance system in strengthening the weak gaze-performance link previously identified in children with ASD during MR-DMT. The results demonstrate that systematically addressing both perceptual engagement and motor transformation processes can significantly enhance the functional utility of visual attention.

\subsection{Discussion}
The findings from Study 2 provide robust empirical support for the efficacy of the proposed dual-level visual guidance system in strengthening the weak gaze-performance link. The results confirm hypothesis 3 that a guidance paradigm specifically designed to scaffold the VMI process can significantly enhance the functional relationship between visual attention and motor execution in children with ASD.

The most compelling evidence comes from the differential effects across experimental conditions. While all guidance conditions improved the gaze-performance correlation compared to baseline, the dual-level guidance condition produced the strongest effects, with the correlation between gaze duration and motor imitation performance increasing from non-significant in the baseline condition to substantial and highly significant. This pattern was consistent in all motor dimensions, suggesting that the combined perceptual and transformational guidance effectively addressed the core VMI deficits.

Notably, the perceptual-only and transformational-only conditions both produced intermediate effects, supporting our theoretical framework that both attention direction and movement translation mechanisms contribute independently to VMI enhancement. Transformational guidance alone produced slightly stronger effects than perceptual guidance alone. This finding suggests that the movement translation component may be especially critical. It appears particularly important for addressing the core imitation difficulties in ASD.

Interestingly, while the correlational link was strongly enhanced, the corresponding increase in absolute performance scores was modest (small effect sizes). This dissociation is highly informative: it suggests that our intervention primarily acted by improving the efficiency of translating visual input into motor output, rather than by universally elevating performance irrespective of gaze behavior.

The replication of Study 1's findings in the baseline condition further strengthens the validity of the weak gaze-performance link as a robust phenomenon in non-guided MR-DMT environments. A consistent dissociation pattern was observed. It showed weaker correlations with motor performance, but stronger associations with target-related responsive behaviors. This pattern reinforces the need for targeted interventions that specifically address VMI processes.

\section{General Discussion}

This research provides a comprehensive investigation into the gaze-performance link in children with ASD during MR-DMT, addressing a critical gap in both theoretical understanding and interventional design. Through two sequential studies, we first identified and then successfully mitigated a fundamental barrier to effective imitation-based learning in this population.

\subsection{Theoretical Contributions and Mechanistic Insights}

The core contribution of this work is the empirical demonstration and subsequent enhancement of the \textit{gaze-performance link}—a crucial but previously overlooked mechanism in MR interventions for ASD. Study 1 systematically established that children with ASD exhibit a significant dissociation between visual perception and motor execution, whereby increased gaze duration on task-relevant areas does not translate to improved motor performance. This finding challenges the implicit assumption underlying many gaze-contingent interventions.

Importantly, the differential pattern of correlations across performance domains provides nuanced insights into the nature of VMI deficits in ASD. While the link between gaze and motor performance was consistently weak, gaze duration showed stronger associations with certain cognitive and engagement measures. This dissociation suggests that the core deficit lies not in attention per se, but in the specific mechanisms that translate visual information into motor plans.

\subsection{Intervention Efficacy and Design Principles}

Building on this foundation, Study 2 demonstrated that a theoretically-guided dual-level visual guidance system can effectively strengthen this compromised link. The differential effectiveness of the perceptual and transformational guidance components provides mechanistic insights into the VMI process. While perceptual guidance successfully directed attention to relevant areas, transformational guidance proved more critical for facilitating motor translation.

The finding that dual-level guidance produced the strongest effects underscores the multi-faceted nature of VMI and supports an integrated approach to intervention design. By simultaneously addressing both \textit{where to look} (perceptual guidance) and \textit{how to translate} (transformational guidance), the system provides the necessary scaffolding to bridge the gap between perception and action.

\subsection{Practical Implications and Clinical Relevance}

From a practical perspective, this research offers a validated framework for developing more effective MR-based interventions for ASD. The dual-level guidance system can be adapted to various educational and therapeutic contexts that require VMI, from motor skill training to social communication programs. The use of affordable and scalable technologies further enhances the practical applicability of this approach.

\subsection{Limitations and Future Directions}

Despite these contributions, several limitations should be acknowledged. The controlled laboratory environment, while necessary for rigorous experimentation, may limit ecological validity. Future research should examine the generalization of these effects to naturalistic settings and longer-term interventions. Additionally, while the sample size was sufficient for detecting main effects, larger samples would enable more sophisticated analyses of individual differences and response heterogeneity.

Looking forward, this work opens several promising directions for future research. Neuroimaging studies could provide valuable insights into the neural mechanisms underlying the observed behavioral changes, particularly in visuomotor integration pathways. Additionally, research exploring the optimal personalization of guidance parameters based on individual cognitive profiles could enhance intervention precision and effectiveness. Further investigations into the transfer of acquired skills to real-world contexts would strengthen the practical significance of these findings, while longitudinal studies examining the maintenance of intervention effects over time would provide important clinical insights for sustainable treatment approaches.

\subsection{Conclusion}

In conclusion, this research makes significant theoretical and practical contributions to the field of ASD intervention. By identifying, quantifying, and successfully addressing the weak gaze-performance link in MR-DMT, we provide both a validated intervention approach and a conceptual framework for understanding visuomotor integration deficits in ASD. The findings demonstrate that through theoretically-guided design, MR technologies can move beyond merely capturing attention to actively facilitating the complex process of translating perception into action—ultimately leading to more effective and engaging interventions for children with ASD.

\acknowledgments{
The authors wish to thank A, B, and C. This work was supported in part by
a grant from XYZ.}

\bibliographystyle{abbrv-doi}

\bibliography{main}
\end{document}